\documentclass[lettersize,journal]{IEEEtran}
\usepackage{amsmath,amsfonts}
\usepackage{algorithm}
\usepackage{algpseudocode}
\usepackage{array}
\usepackage{textcomp}
\usepackage{stfloats}
\usepackage{url}
\usepackage{verbatim}
\usepackage{graphicx}
\usepackage{enumitem}
\usepackage{eurosym}
\usepackage{tabularray}
\usepackage{graphicx}
\usepackage{hyperref}
\hypersetup{
    colorlinks = true,
    citecolor = blue,
    linkcolor = blue,
    urlcolor = blue,
    bookmarksopen = true,
    filecolor=blue
}

\usepackage{booktabs}
\usepackage{multirow}
\usepackage{balance}

\usepackage{color}

\begin{document}
%
\title{Optimizing Electric Vehicles Charging using Large Language Models and Graph Neural Networks}

\author{Stavros Orfanoudakis~\IEEEmembership{Student~Member,~IEEE}, 
Peter~Palensky,~\IEEEmembership{Senior~Member,~IEEE}, Pedro~P.~Vergara,~\IEEEmembership{Senior~Member,~IEEE}
\thanks{This study is funded by the HORIZON Europe Drive2X Project 101056934. All the authors are with the Intelligent Electrical Power Grids (IEPG) Section, Faculty of Electrical Engineering, Mathematics and Computer Science, Delft University of Technology, Delft, The Netherlands (emails: {s.orfanoudakis, p.palensky, p.p.vergarabarrios}@tudelft.nl)}}
\maketitle

\begin{abstract}
Maintaining grid stability amid widespread electric vehicle (EV) adoption is vital for sustainable transportation. Traditional optimization methods and Reinforcement Learning (RL) approaches often struggle with the high dimensionality and dynamic nature of real-time EV charging, leading to sub-optimal solutions. To address these challenges, this study demonstrates that combining Large Language Models (LLMs), for sequence modeling, with Graph Neural Networks (GNNs), for relational information extraction, not only outperforms conventional EV smart charging methods, but also paves the way for entirely new research directions and innovative solutions.
\end{abstract}

\begin{IEEEkeywords}
EV Charging, Large Language Models (LLMs), Reinforcement Learning (RL), Graph Neural Networks (GNNs)
\end{IEEEkeywords}

\IEEEpeerreviewmaketitle

\section{Introduction}

\IEEEPARstart{T}{he}  efficient utilization of electric vehicle (EV) charging infrastructure is critical in leveraging existing energy resources effectively and preventing additional congestion on electricity grids. For instance, congestion in major markets incurres billions in additional costs annually~\cite {Doying2023}, highlighting the urgent need for scalable solutions that ensure the seamless integration of EVs into distribution systems. Addressing these challenges is essential not only for maintaining grid stability but also for supporting the widespread adoption of EVs, thereby contributing to sustainable transportation.

Despite the progress made with traditional optimization and Reinforcement Learning (RL), current methods often fail to manage the complexities and scalability required by modern EV charging demands. Traditional techniques, such as mathematical programming and model predictive control, often struggle with the high dimensionality and dynamic nature of real-time charging scenarios~\cite{7500060}, leading to inefficiencies and delayed decision-making. Similarly, existing RL frameworks, including multi-agent systems, face challenges in ensuring constraint satisfaction and generalizing across diverse environments.
To overcome these limitations,
in this paper, we demonstrate how recent advances in powerful Large Language Models (LLMs) and Graph Neural Networks (GNNs) can be used to achieve accurate EV smart charging.
To the best of our knowledge, this paper is the first to enable LLMs for decision-making problems related to EV charging. Extensions of the proposed model to other problems pave the way for entirely new research directions and innovative solutions.

In this letter, three main contributions are presented. Firstly, we introduce an innovative embedding that reinterprets the EV charging problem as a graph representation, thereby enabling the use of powerful GNNs to capture the intricate relational dynamics of charging infrastructure. We are the first to integrate LLMs for EV charging decision-making, setting a new paradigm in the application of LLMs to this domain. Finally, we extensively demonstrate that the synergy between powerful LLMs and dynamic GNNs substantially enhances the accuracy and reliability of smart charging.

\section{EV Smart Charging}
From the perspective of a Charge Point Operator (CPO), the Vehicle-to-Grid (V2G) smart charging problem involves strategically managing the charging and discharging of EVs to maximize profits while adhering to critical operational constraints. The primary objective for CPOs is to capitalize on dynamic electricity pricing by charging EVs when energy prices are low and discharging back to the grid during periods of higher prices, thereby generating revenue through arbitrage.
This profit maximization is balanced by the need to comply with aggregated grid power constraints, ensuring that the total power drawn or supplied by all connected EVs does not exceed the limits set by the distribution system operator.
Additionally, CPOs must guarantee that each EV receives enough energy by their departure time, necessitating precise scheduling. The network and user constraints make the V2G smart charging a complex challenge that requires advanced decision-making methods to achieve high-quality results in real-time.

\begin{figure*}[!t]
\centering
\includegraphics[width=1\linewidth]{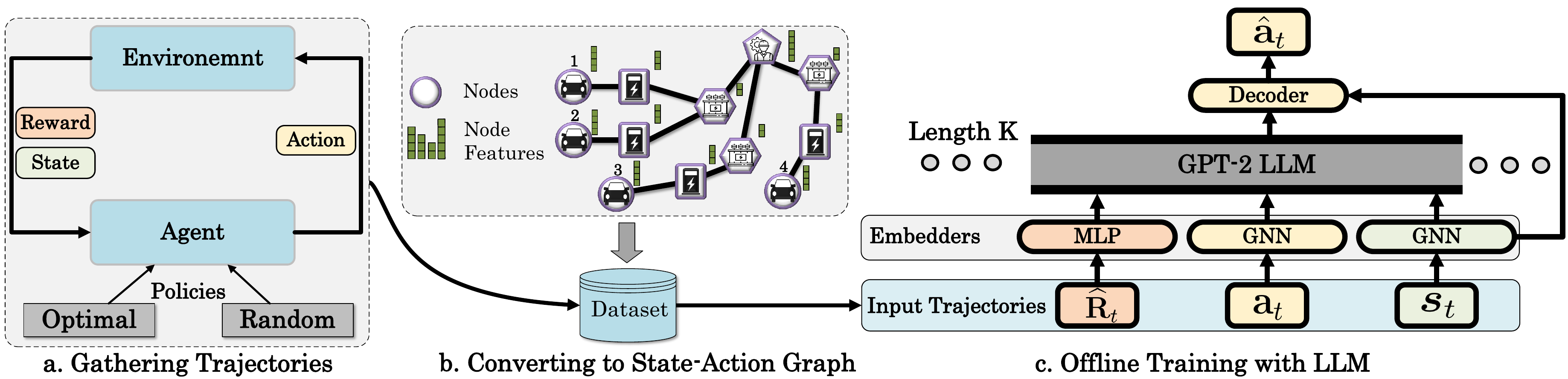}
\caption{Overview of the proposed learning method. Trajectories are first generated by interacting with the environment using optimal and random policies. The resulting state and action vectors are then transformed into graph representations to efficiently capture relational dependencies and support dynamic environments. Finally, an offline RL method, powered by an LLM, is trained to optimize charging strategies and maximize future rewards.
}
\label{fig:framework}
\end{figure*}

\section{LLMs for Smart Charging}

Using LLMs for EV smart charging consists of three steps, as illustrated in Fig.~\ref{fig:framework}: (a) collecting trajectories, (b) converting state-action representations from vectors to graphs, and (c) offline RL training using an LLM as a core.

\subsection{Data Gathering for Offline RL}

Offline RL is employed to develop optimal policies using a fixed dataset of historical interactions, thereby avoiding the risks and costs associated with additional environment exploration. In the context of EV charging, offline RL leverages a dataset \( \mathcal{D} = \{(s, a, r)\} \) derived from past charging sessions optimally managed by CPOs. To further enhance exploration, trajectories generated under a random uniform policy are incorporated (Fig.~\ref{fig:framework}.a). These trajectories capture the states \( s_t \), actions \( a_t \), and rewards \( r_t \) experienced under diverse conditions, such as variable electricity prices, different EVs arrival and departure times, and grid constraints. The goal is to learn a policy \( \pi_\theta(a \mid s) \) that maximizes the expected discounted return $\mathbb{E}\left[\sum_{t=0}^{\infty} \gamma^t R(s_t, a_t)\right]$, where \( \gamma\) is the discount factor. By training on this static dataset, offline RL models can uncover effective charging and discharging strategies that minimize costs and maintain grid stability.

\subsection{GNN Embeddings}

Directly using the gathered trajectories is inefficient because raw, unprocessed data limits a neural network's ability to learn effectively.
To improve the learning capabilities of the LLM, the EV charging problem is reinterpreted as a graph where each node represents an entity, such as an EV, a charger, a transformer, or the CPO, and the edges capture the interactions (power or information exchange) between these entities (Fig.~\ref{fig:framework}.b). 
This graph-based representation naturally incorporates both local and global constraints in the system in a more efficient way than having a long state vector.
Each node in the graph has a set of features that are initially processed by dedicated Multi-Layer Perceptrons (MLPs) to generate informative embeddings. These embeddings are further refined through successive graph convolution layers, which message-pass among neighboring nodes to aggregate and combine contextual information. Consequently, the final embeddings, used as input for the LLM (see Fig.~\ref{fig:framework}.c), capture both the local context of individual nodes and the overall structure of the charging network. 
Integrating these rich, graph-based representations~\cite{Orfanoudakis2024} into the RL framework enables the model to make more informed decisions regarding charging and discharging actions, surpassing traditional vector representations.

\subsection{Learning with LLMs}  
After generating a diverse trajectory dataset using both optimal and random policies, we train an LLM to maximize the expected reward. To achieve this, we employ the Decision Transformer (DT) algorithm~\cite{chen2021decision}, which is an LLM-based model that reframes RL as a sequence prediction task. By leveraging the causal-attention mechanism, the LLM captures dependencies across time steps, enabling it to model long-term relationships in the data.
The DT is trained on the gathered processed trajectories dataset as shown in Fig.~\ref{fig:framework}.c.
During training, the model predicts the next action in each sequence and adjusts its parameters by comparing these predictions to the actions actually taken. This approach allows the model to identify which charging and discharging actions yield the best outcomes, gradually adapting to complex temporal patterns, such as energy costs and grid performance over time.
The GNN embedders process the underlying graph structure of the charging network to generate rich contextual embeddings, which are then fed into the DT. This additional context helps the model make more informed decisions improving the overall learning performance.

\section{Results}

\begin{table*}[t]
\centering
\caption{Comparison of key metrics for a problem with 50 charging stations problem after 100 simulation episodes.}
\label{tab:rewards_data}
\resizebox{1\textwidth}{!}{
\begin{tblr}{
  cells = {c},
  hline{1,7} = {-}{0.08em},
  hline{2} = {-}{},
}
Algorithm & {Energy \\Charged [MWh]} & {Energy \\Discharged [MWh]} & {User \\Sat. [\%]} & {Min. User \\Sat. [\%]} & {Power \\Violation [kW]} & Costs [€] & Reward [$\times10^5$] & {Step time\\~[sec/step]}\\
CAFAP & $2.2$ ±$0.2$ & $0.00$ ±$0.00$ & $100.0$ ±$0.0$ & $100.0$ & $3447.0$ ±$447.2$ & $-462$ ±$252$ & $-4.581$ ±$0.489$ & $0.001$\\
Business as Usual & $2.0$ ±$0.2$ & $0.00$ ±$0.00$ & $98.9$ ±$0.4$ & $59.5$ & $9.5$ ±$6.5$ & $-382$ ±$215$ & $-0.961$ ±$0.080$ & $0.001$\\
PPO (Online RL) & $1.3$ ±$0.1$ & $0.07$ ±$0.02$ & $85.6$ ±$1.8$ & $16.2$ & $168.9$ ±$140.5$ & $-237$ ±$130$ & $-2.049$ ±$0.311$ & $0.002$\\
GNN+LLM (\textbf{Ours}) & $1.5$ ±$0.2$ & $0.28$ ±$0.04$ & $97.5$ ±$0.5$ & $70.1$ & $64.2$ ±$67.1$ & $-227$ ±$131$ & $-0.096$ ±$0.073$ & $0.012$\\
Optimal (Oracle) & $3.1$ ±$0.3$ & $1.77$ ±$0.29$ & $99.0$ ±$0.2$ & $76.1$ & $10.9$ ±$14.7$ & $-205$ ±$126$ & $-0.032$ ±$0.025$ & $-$
\end{tblr}
}
\end{table*}

In this section, we present experimental results illustrating the superior performance of the proposed approach. The simulations are conducted in EV2Gym~\cite{10803908}, a simulator that leverages realistic EV behavior data, thereby increasing confidence that the outcomes generalize to real‐world settings. We focus on a challenging use case with 50 charging stations, where EVs arrive without notice and the distribution system operator practices dynamic aggregated power limits. This scenario poses a complex, large‐scale problem that baseline methods struggle to tackle effectively. To demonstrate our method’s advantages, we compare it against multiple algorithms. The heuristic baselines include the “Charge As Fast As Possible” (\textit{CAFAP}) strategy, which delivers energy immediately, and a cyclical priority round‐robin approach commonly referred to as “Business as Usual” (\textit{BaU}). We also evaluate state‐of‐the‐art online RL algorithms, where the best (\textit{PPO}) was selected for comparison. For completeness, an offline \textit{Optimal} policy algorithm is included that, even if impossible in practice, solves the problem with perfect knowledge of future events.

\begin{figure}[!t]
\centering
\includegraphics[width=0.9\linewidth]{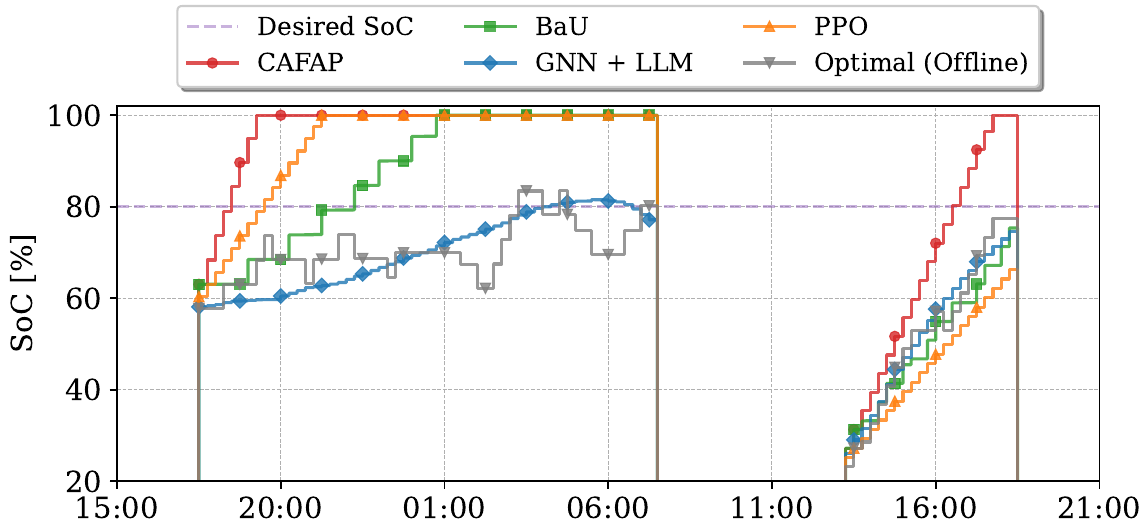}
\includegraphics[width=1\linewidth]{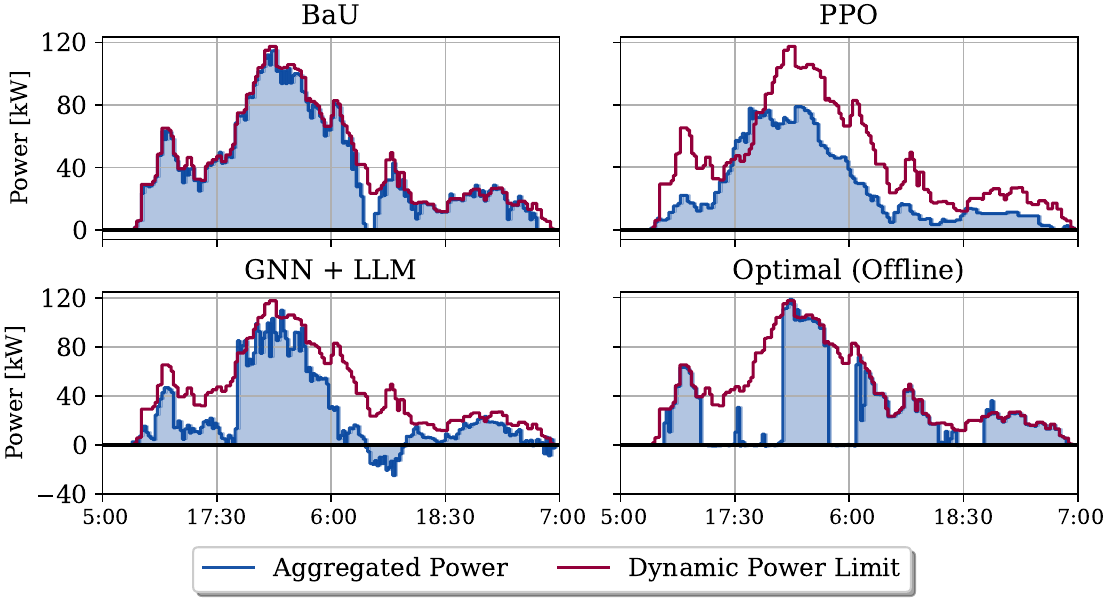}
\caption{Visual comparison of a use case with 50 chargers showing: a) an individual charger's charging profile w.r.t. the connected EVs, and b) the aggregated power following the grid operator's power limit.
}
\label{fig:use_case}
\end{figure}

Fig.~\ref{fig:use_case} compares the performance of baseline algorithms with the proposed LLM-based method for addressing the EV smart charging problem in a scenario featuring 50 charging stations and dynamic grid power constraints. As shown in Fig.~\ref{fig:use_case}.a, baseline methods struggle to learn the complex underlying dynamics, whereas our \textit{GNN+LLM} approach consistently charges EVs to the desired capacity while capitalizing on V2G opportunities generating charging schedules close to the \textit{Optimal}. Similarly, Fig.\ref{fig:use_case}.b demonstrates that the \textit{GNN+LLM} method effectively maintains power consumption within the grid’s dynamic limits.

Table~\ref{tab:rewards_data} presents a detailed comparison after 100 evaluations.
Compared to the heuristic baselines (\textit{CAFAP} and \textit{BaU}), the \textit{GNN+LLM} achieves a much better trade‐off between user satisfaction, costs, and grid power violations. For example, \textit{CAFAP} perfectly satisfies all users (100\% user satisfaction) but incurs very high power violations (over 3,400 kW) and ends up with large negative costs. On the other hand, \textit{BaU} keeps power violations low but exhibits a significant drop in minimum user satisfaction (only 59.5\%), while also incurring higher costs than our method. Looking at RL baselines, the \textit{PPO} algorithm has worse user satisfaction on average (85.6\%) and significantly larger power violations (168.9 kW), suggesting it fails to balance the competing objectives of cost reduction and user comfort. In contrast, the proposed method stands out by keeping both average and minimum user satisfaction close to the optimal, while also reducing costs and limiting grid violations to moderate levels. Our method demonstrates a strong balance across all key metrics in real time outperforming heuristic and online RL strategies.

Table~\ref{tab:reward_breakdown} compares the performance of the proposed LLM-based approach when trained on three different datasets: purely Random, purely Optimal, and a combined (Optimal + Random) dataset. The second column shows the number of trajectories sampled for each training set, while the third column reports the average reward observed in the dataset. The fourth column provides the maximum reward that the trained model achieves. Notably, although the combined dataset has a lower average reward than the Optimal‐only dataset, the resulting trained policy attains the highest max reward. This highlights how incorporating diverse trajectories, including suboptimal ones, helps the LLM learn a more robust and effective policy overall enabling effective EV smart charging.

\begin{table}
\centering
\caption{Maximum reward of GNN+LLM trained on merged Optimal and Random datasets, with bold showing the best dataset.}
\label{tab:reward_breakdown}
\begin{tblr}{
  cells = {c},
  hline{1,5} = {-}{0.08em},
  hline{2} = {-}{},
}
{Training\\Dataset} & {Samples\\Trajetories} & {Average Dataset\\~Reward $[\times10^{5}]$} & {Max Reward\\ $[\times10^{5}]$ }\\
Random & 5000 & $-3.79$ ±$0.58$ & $-2.13$\\
Optimal & 5000 & $-0.03$ ±$0.02$ & $-0.25$\\
Optimal + Random & 2500 + 2500 & $-1.90$ ±$1.97$  & $\mathbf{-0.07}$
\end{tblr}
\end{table}

\section{Conclusions}
In this study, we have demonstrated that integrating LLMs with GNNs significantly enhances charging schedule optimization for complex, large-scale EV charging scenarios. Our robust experimental evidence confirms that this hybrid approach outperforms traditional RL algorithms by effectively capturing long-term dependencies and intricate network relationships, thereby addressing the scalability and dynamic challenges that conventional methods struggle to overcome.


\bibliographystyle{IEEEtran}
\bibliography{ref}

\end{document}